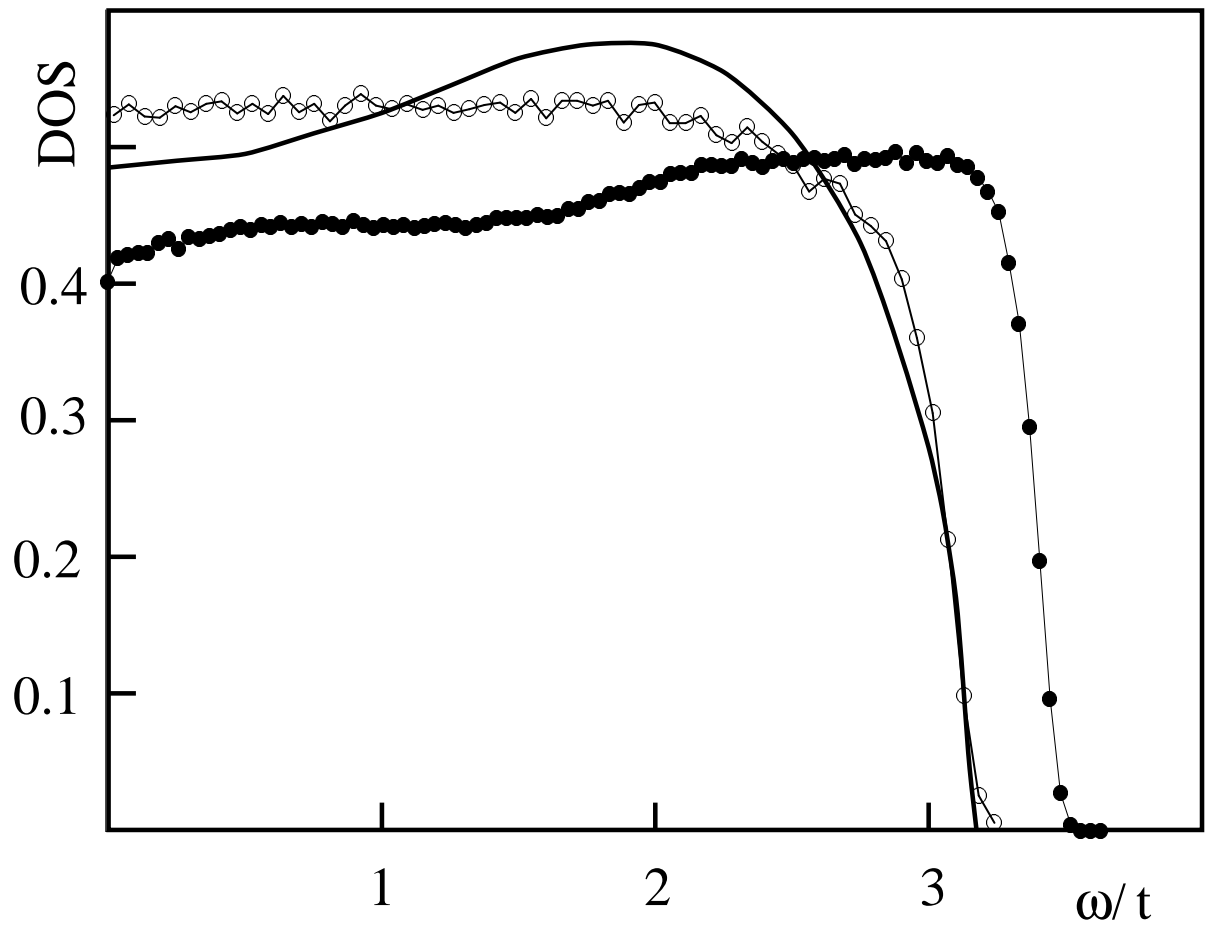

# Mean–field limit of the random flux model

B. Kramer

*I. Institut für Theoretische Physik, Universität Hamburg, Jungiusstraße 9, D-20355 Hamburg, Germany*

D. Belitz

*Department of Physics and Materials Science Institute, University of Oregon, Eugene, OR 97403, USA*

M. Batsch

*Physikalisch–Technische Bundesanstalt Braunschweig, Bundesallee 100, D-38116 Braunschweig, Germany*

*and*

*I. Institut für Theoretische Physik, Universität Hamburg, Jungiusstraße 9, D-20355 Hamburg, Germany*

(July 5, 1996)

The problem of non–interacting electrons on a square lattice subject to a random magnetic flux is mapped onto a one–dimensional model with infinitely many orbitals per site. Linking each orbital with $N(\gg 1)$ other orbitals maps the problem onto Wegner's $N$-orbital model in the same limit, while the original problem corresponds to $N = 1$. The exact solution for $N = \infty$, the mean–field limit, is discussed and compared with numerical results. An outline of a $1/N$-expansion is given.

The so-called random flux model (RFM) of electrons on a square lattice with a random magnetic flux penetrating each plaquette has been the subject of some debate lately. Of particular interest is the question whether or not there are extended states and/or localization transitions in such a system. Some investigations of this model have been motivated by the quantum Hall problem at half filling: After a transformation to composite fermions, the fermionic quasiparticles at half filling are subject only to a fluctuating vector potential, but not to a mean magnetic field.[1] In the presence of disorder, the random flux model is believed to provide a good representation of this situation, due to a constraint in the composite fermion theory that ties the fluctuation of the vector potential to those of the fermion number density.[2] Similar models of fermions with a gauge interaction have been proposed in the context of high-$T_c$ superconductivity.[3] Finally, an electron moving in a random flux also provides an interesting localization problem in its own right.

No definite conclusion about the nature of the states at the band center was obtained from various numerical investigations of the RFM. They were based on studies of the participation ratio, the conductance distribution, the scaling of transmission properties, quantum diffusion and Chern numbers. The existence of extended states near the band center was suggested.[4–7] However, data that are consistent with complete localization,[8,9] or with a whole region of critical states[10] were also obtained. Analytically, Aronov et al.[11] have shown that perturbatively the electron dynamics is diffusive. From this, they drew the conclusion that the model is in the same universality class as a supersymmetric nonlinear sigma–model with unitary symmetry, which is known to have only localized states in two dimensions (2D). However, the work of Zhang and Arovas[12] suggests that this argument may be misleading, for a similar reason for which its analog in the case of a homogeneous magnetic field fails. These authors argue that a term describing the fluctuations of a topological density must be added to the nonlinear sigma–model, and that the system exhibits a Kosterlitz–Thouless transition from a region with power–law localization at the band center to one with exponential localization in the band tails.

In this paper, we propose an alternative approach to the RFM. We show that by means of a gauge transformation[13] the model can be mapped onto a one–dimensional (1D) model with infinitely many orbitals per site that is reminiscent of Wegner's $N$–orbital model.[14] A modification that links each orbital to $2N$ other orbitals provides, in the mean–field limit $N \to \infty$, an exact mapping onto Wegner's model in the same limit, where a solution is known.[14] The original RFM corresponds to $N = 1$. The conductivity in the mean–field limit is non-zero everywhere in the band. While this is certainly qualitatively different from the behavior at $N = 1$, we expect the density of states (DOS) to be less strongly dependent on $N$. Indeed, our mean–field result for the DOS agrees reasonably well with numerical results for $N = 1$. We also outline a $1/N$-expansion that will provide a new way to obtain information about the transport properties of the RFM.

We start with the tight binding representation of the RFM,

$$H = -t \sum_{\langle \mathbf{x},\mathbf{y} \rangle} e^{i\theta_{\mathbf{x},\mathbf{y}}} |\mathbf{x}\rangle \langle \mathbf{y}| \quad . \tag{1}$$

Here the $|\mathbf{x}\rangle$ are electron states on the sites $\mathbf{x}$ of a square lattice with unit lattice constant. The sum is over nearest neighbors only, and $t > 0$ is the hopping matrix element. The phases $\theta_{\mathbf{x},\mathbf{y}}$ are independent random variables apart from the hermiticity requirement $\theta_{\mathbf{x},\mathbf{y}} = -\theta_{\mathbf{y},\mathbf{x}}$. We assume that they are independently distributed within the interval $[-\pi, \pi]$. Next we recall that the phase disorder can be restricted to 1D by means of a gauge transformation.[13] Let $\mathbf{e}_j$ be the unit vector in $j$-direction ($j = 1, 2$), and define $\theta_{\mathbf{x}, \mathbf{x}+\mathbf{e}_j} \equiv \theta_{\mathbf{x}}^{(j)}$. Now consider the states $|\mathbf{x}\rangle = |x_1, x_2\rangle \equiv |r, m\rangle$. From now on we assume the system to have a length $R$ in the $r$–direction, with periodic boundary conditions. In the $m$–direction we assume a finite width $M$. A local gauge transformation

$$|r, m\rangle \rightarrow e^{i\varphi_{r,m}} |r, m\rangle \quad , \tag{2}$$

with $\varphi_{r,m} \equiv \sum_{n=1}^{m-1} \theta_{r,n}^{(2)}$ transforms the Hamiltonian into

$$H = -\sum_{r,r'} \sum_{m,m'} \left[ v_{r,r'} \delta_{m,m'} + \frac{1}{\sqrt{M}} f_{r,r'}^{m,m'} \right] |r, m\rangle \langle r', m'|. \tag{3a}$$

Here

$$v_{r,r'} = t \left[ \delta_{r',r+1} + \delta_{r',r-1} \right], \tag{3b}$$

and

$$f_{r,r'}^{m,m'} = t \delta_{r,r'} \sqrt{M} \left[ \delta_{m',m+1} e^{i\phi_{r,m}} + \delta_{m',m-1} e^{-i\phi_{r,m-1}} \right]. \tag{3c}$$

The $\phi_{r,m}$ are linear combinations of the $\theta_{r,m}^{(1,2)}$,

$$\phi_m(r) \equiv \phi_{r,m} = \theta_{r,m}^{(1)} + \varphi_{r,m} - \varphi_{r+1,m}. \tag{4}$$

Apart from hermiticity, they are therefore also independent random variables that obey the same distribution as the $\theta_{r,m}^{(1,2)}$. In particular,

$$\langle e^{i\phi_m(r)} e^{-i\phi_{m'}(r')} \rangle = \delta_{r,r'} \delta_{m,m'}, \tag{5a}$$

$$\langle e^{i\phi_m(r)} e^{i\phi_{r'}(r')} \rangle = 0. \tag{5b}$$

The gauge transformation has eliminated the disorder in the $r$–direction. In the absence of the disorder in the Hamiltonian, i.e. the random variables $f_{r,r'}^{m,m'}$ in Eq. (3a), we would simply have a 1D tight binding model that can be diagonalized by means of a Fourier transform from sites $r$ to wavenumbers $q$. The corresponding Green's function is[15]

$$G^{(0)}(q, z) = [z - 2t \cos q]^{-1}, \tag{6}$$

with $z = \omega + i0$, the complex frequency or energy variable. With the random matrix elements $f_{r,r'}^{m,m'}$, the model written in the form of Eqs. (3) can be interpreted as describing a 1D system with $M$ orbitals per site $r$. We are interested in the behavior in the 2D limit $R, M \to \infty$.

The model as defined by Eqs. (3) is formally very similar to Wegner's site–diagonal $N$–orbital model[14] (with $N \equiv M$). We have deliberately chosen our notation so as to make this obvious. A crucial difference, however, is that in the $N$-orbital model, Wegner assumed all of the $f_{r,r'}^{m,m'}$ with $m \neq m'$ to be equally distributed, while in our case $f$ is nonzero only if $m$ and $m'$ are nearest neighbors. Wegner's results therefore do *not* directly apply to the present case.

In order to proceed, let us redefine the random variables $f$, Eq. (3c),

$$f_{rr'}^{m,m'} \equiv t \, \delta_{r,r'} \, g_{m-m'} \, e^{i\phi_{m,m'}(r)}, \tag{3c'}$$

where the $\phi_{m,m'}(r)$ are independent random phases with equal probability distributions for all $m \neq m'$ that link points $(r, m)$ and $(r, m')$, and the $g_{m-m'}$ determine the range of $f$ in $m$-direction. The original random flux model corresponds to $g_m \equiv g_m^{(1)} = \sqrt{M}(\delta_{m,1} + \delta_{m,-1})$ in the limit $M \to \infty$.

We now define a new model by choosing $g_m \equiv 1$. Then, the Hamiltonian is given by Eqs. (3), however with random variables

$$f_{r,r'}^{m,m'} \equiv t \, \delta_{r,r'} \, e^{i\phi_{m,m'}(r)}, \tag{7a}$$

that are independently distributed with a second moment,

$$\langle f_{r,r'}^{m,m'} f_{s,s'}^{n,n'} \rangle = t^2 \delta_{r,r'} \, \delta_{r,s} \, \delta_{s,s'} \, \delta_{m,n'} \, \delta_{m',n}. \tag{7b}$$

Equations (3) and (7) now desribe Wegner's unitary, site–diagonal model.[14] This has been achieved by means of the mean–field construction of linking each orbital to every other orbital on the same site.

For the model described by Eqs. (3) and (7) in the limit $M \to \infty$, an exact solution is known for both the single–particle Green's function, and in particular for the density of states, and for the transport properties.[14] This is due to the fact that, as a consequence of Eqs. (7), all terms in the perturbation expansion for the Green's function that contains crossed impurity lines vanish in the limit $1/M \to 0$. The remaining diagrams can be summed exactly.

The resulting Green's function can be written,

$$G(q, z) = [z - 2t \cos q + \Sigma(z)]^{-1}, \tag{8}$$

where $\Sigma$ is the self energy. The exactly solvable limit mentioned above corresponds to the self–consistent Born approximation for the self energy becoming exact, i.e. $\Sigma$ is given by,[14]

$$\Sigma(z) = t^2 \sum_q G^{(0)}(q, z + \Sigma(z)), \tag{9}$$

with $G^0(q, z)$ from Eq. (6). By performing the $q$-integral we obtain a quartic equation for the retarded self energy, $\Sigma_R(\omega) = \Sigma(z = \omega + i0)$,

$$\Sigma_R(\omega) = \frac{-1}{\sqrt{(\omega + \Sigma_R(\omega))^2 - 4} + i0}. \quad (10a)$$

Here we have assumed $t = 1$, i.e. we measure all energies in units of half of the 1D perfect crystal bandwidth. Equation (10a) is easily solved, and from its solution we obtain the DOS $N(\omega)$ as

$$N(\omega) = \frac{1}{\pi} \text{Im} \Sigma_R(\omega). \quad (10b)$$

Figure 1 shows our mean–field result, Eqs. (10), for the DOS together with the result of numerical calculations for the model with $g_m = g_m^{(1)}$ (see the definition after Eq. (3c')), and for models with additional couplings between orbitals $(m, m \pm 2)$ and a given site $r$ of the 1D chain corresponding to $N = 2$ (see Eq. (11) below). The latter were obtained by numerically diagonalizing systems with the sizes $40^2$, $80^2$, $120^2$ ($N = 1$) and $40^2$ ($N = 2$). The accuracy of the results is better than 3% ($N = 1$) roughly 10% ($N = 2$).

The Lifshitz–bounds of the spectrum for $N = 1$ (cf. Eq. (1)) are $\pm 4$. From Eqs. (7) we see that the disorder induced by the random phases is $t$, as large as $1/8$ of the unperturbed bandwidth. This strong disorder is reflected in the DOS: only a weak maximum in the vicinity of $\omega = 2t$ is left over of the square root divergence in the unperturbed DOS at the perfect 1D crystal band edge, and the band edge is shifted by more than 50%. Lifshitz tails are, as usual, suppressed in our mean–field model. However, as can be seen from the numerical data in Fig. 1, the tails are extremely small, even in the orginal model, and the overall qualitative features of the numerical result are well reproduced by the mean–field DOS.

There are, however, also important differences between the mean–field result and the exact numerical DOS: In the numerical result for $N = 1$, the total bandwidth is about 10% larger than in the mean–field result. This can be qualitatively understood by the following argument. For $g_m = g_m^{(1)}$, and no phase disorder, the band edge would be at $\omega = 4t$, and the DOS would have a van Hove singularity in the center of the band. In the presence of phase disorder, the latter is smeared out due to the absence of translational invariance. Simultaneously, the width of the band is reduced, since the average kinetic energy corresponding to the $m$–direction decreases as a consequence of the randomization of the phases. In the mean–field model, the phases are effectively even more randomized so that the band width is further diminished. This argument is supported by first numerical results for models with an increased number of couplings in the $m$–direction. Already when next–nearest neighbors are taken into account (Fig. 1), the width of the band is reduced to approximately the value predicted by the mean–field theory with $N = \infty$. (In the center of the band convergence with respect to the system size is not reached for the data shown in Fig. 1, were for $N = 2$ only systems with sizes up to $40^2$ have been considered).

The net result of the strong phase disorder is a density of states that bears no similarity any more with that of a two–dimensional ordered system, but rather resembles that of a *one–dimensional* system with a very strong rounding of the van Hove singularities at the band edges. As can be seen from the figure, this qualitative feature is well captured by our mean–field theory.

The conductivity in the mean–field limit is simply given by the square of the DOS.[14] This means that the conductivity is non-zero everywhere within the band, and hence all states are extended. This feature is certainly characteristic of the mean–field character of the model. Information about the transport properties of the model with $g_m = g_m^{(1)}$ can therefore only be ontained by calculating corrections to the mean–field limit.

In order to set up a systematic perturbation theory with the mean–field model as zeroth approximation, we define the function $g_m$ in Eq. (3c') as

$$g_m = \begin{cases} \sqrt{M/N} & \text{if } 0 < |m| \leq N \\ 0 & \text{else} \end{cases} \quad (11)$$

This choice couples $2N$ orbitals on each site. In the limit $M \to \infty$, $N = M$ yields the mean–field model solved above, while $N = 1$ yields the original model as defined in Eqs. (3). It is now possible to set up a systematic expansion in powers of $1/N$, along the lines of Ref. 16. While this should yield only small, quantitative corrections to the DOS, we expect an expansion of the conductivity in powers of the small parameter $1/N$ to show indications of localization, starting at $O(1/N^2)$. Such a calculation will be rather different from the perturbation theory of Ref. 11, and especially the dependence of the results on the location within the band should be very interesting. The fact the the above numerical DOS with $N = 2$ agrees already quite well with the mean–field result suggests also that very probably the expansion with respect to $1/N$ will converge quite rapidly. The results of such an approach will be reported elsewhere.

In summary, we have constructed a mean–field limit for the random flux model. We have shown that the mean–field DOS reproduces the qualitative features of exact numerical results. A systematic method to expand about the mean–field limit has also been proposed.

This work was supported by NATO (grant No. CRG–941250), by the NSF (grant No. DMR–95–10185), by the Science and HCM programmes of the EU (contracts SCC-CT90–0020, CHRX-CT93–0126), by the Deutsche Forschungsgemeinschaft via the Graduiertenkolleg "Physik nanostrukturierter Festkörper" and Projekt No. KR 627/8-1.

[1] B. I. Halperin, P. A. Lee, and N. Read, Phys. Rev. B **47**,

7312 (1993).

[2] Actually, the constraint ties the fluctuating *field* to the number density, rather than the vector potential. This fact has prompted the consideration of a random field model in Ref. 11. However, there are no indications that considering a random flux instead of a random field makes a qualitative difference.

[3] L. B. Ioffe and A. I. Larkin, Phys. Rev. B **39**, 8988 (1989); N. Nagaosa and P. A. Lee, Phys. Rev. Lett. **64**, 2450 (1992).

[4] C. Pryor, A. Zee, Phys. Rev. B **46**, 3116 (1992).

[5] Y. Avishai, Y. Hatsugai, M. Kohmoto, Phys. Rev. B **47**, 9561 (1993).

[6] V. Kalmeyer, S.–C. Zhang; Phys. Rev. B **46**, 9889 (1992).

[7] D. N. Sheng, Z. Y. Weng, Phys. Rev. Lett. **75**, 2388 (1995)

[8] T. Sugiyama, N. Nagaosa, Phys. Rev. Lett. **70**, 1980 (1993).

[9] D. K. K. Lee, J. T. Chalker, Phys. Rev. Lett. **72**, 1510 (1994).

[10] T. Kawarabayashi, T. Ohtsuki, Phys. Rev. B**51**, 10987 (1995)

[11] A. G. Aronov, A. D. Mirlin, and P. Wölfle, Phys. Rev. B **49**, 16609 (1994).

[12] S.–C. Zhang and D. Arovas, Phys. Rev. Lett. **72**, 1886 (1994).

[13] T. Ohtsuki, Y. Ono, and B. Kramer, J. Phys. Soc. Japan **63**, 685 (1994).

[14] F. Wegner, Phys. Rev. B **19**, 783 (1979).

[15] see, e.g., E. N. Economou, *Green's functions in quantum physics*, Springer (New York 1979).

[16] R. Oppermann and F. Wegner, Z. Phys. B **34**, 327 (1979); R. Oppermann and K. Jüngling, Z. Phys. B **38**, 93 (1980).

FIG. 1. The density of states at $N = \infty$ (solid line) compared to the numerical results for $N = 1$ (●) and $N = 2$ (○). The accuracy of the numerical data is 3% and about 10% for $N = 1$ and $N = 2$, respectively.

4